**Interdot Lead Halide Excess Management in PbS Quantum Dot Solar Cells**

*Miguel Albaladejo-Siguan, David Becker-Koch, Elizabeth C. Baird, Yvonne J. Hofstetter, Ben P. Carwithen, Anton Kirch, Sebastian Reineke, Artem A. Bakulin, Fabian Paulus and Yana Vaynzof\**

M. Albaladejo-Siguan, D. Becker-Koch, E.C. Baird, Y. Hofstetter, A. Kirch, Prof. S. Reineke, Dr. F. Paulus, Prof. Y. Vaynzof
Integrated Center for Applied Physics and Photonic Materials and Center for Advancing Electronics Dresden (cfaed)
Technische Universität Dresden
Nöthnitzer Str. 61, 01187 Dresden, Germany

B. P. Carwithen, Dr. A. A. Bakulin
Department of Chemistry and Centre for Processable Electronics, Imperial College London, London, UK

E-mail: yana.vaynzof@tu-dresden.de

Light-harvesting devices made from PbS quantum dot (QD) absorbers are one of the many promising technologies of third-generation photovoltaics. Their simple, solution-based fabrication together with a highly tunable and broad light absorption makes their application in newly developed solar cells particularly promising. In order to yield devices with reduced voltage and current losses, PbS QDs need to have strategically passivated surfaces, most commonly achieved through lead iodide and bromide passivation. The interdot spacing is then predominantly filled with residual amorphous lead halide species that remain from the ligand exchange, thus hindering efficient charge transport and reducing device stability. Herein, we demonstrate that a post-treatment by iodide based 2-phenylethlyammonium salts (X-PEAI) and intermediate 2D perovskite formation can be used to manage the lead halide excess in the PbS QD active layer. This treatment results in improved device performance and increased shelf-life stability, demonstrating the importance of interdot spacing management in PbS quantum dot photovoltaics.

## 1. Introduction

Photovoltaic diodes fabricated from lead sulfide (PbS) quantum dots (QDs) have piqued the interest of researchers for over a decade,[1–3] resulting in a particularly impressive increase in performance in recent years.[4–6] PbS is well-suited for photovoltaic applications due to its high absorption in the visible and near-infrared wavelengths, the latter being particularly important since they represent an essential constituent of the long-tailed solar irradiance spectrum.[7,8] PbS is especially useful in its quantum dot form, which allows tuning its optoelectronic properties by simply changing the QDs' size, shape and surface composition.[9,10] For single junction solar cells, the QD size is selected based on the optimal 1.3 eV bandgap value ($E_g$), predicted by the thermodynamic theory,[11] but other applications such as tandem cells or infrared detectors benefit from the possibility to tune the PbS QD bandgap from its bulk value of $E_g = 0.37$ eV up to the desired range.[12–16]

Small nanocrystals of PbS ($E_g$ = 1.0-1.5 eV) can be conveniently grown in solution, keeping the QDs in an easy-to-handle dispersion throughout the fabrication process. This ink-like character of PbS QD solution simplifies the fabrication of devices, opening the route for their large scale deposition by spraying or inkjet printing.[14,17,18] Due to their strong and spectrally extended absorption of light, PbS QD solar cells were shown to lead to short-circuit currents ($J_{SC}$) over 30 mA cm$^{-2}$ and fill factors (FF) as high as 70%. These high photovoltaic performance parameters result in power conversion efficiencies (PCE) of up to 14%,[19] which can be improved further by minimizing voltage losses in PbS QD devices.

Since Pb chalcogenide quantum dots have an excess of surface Pb atoms,[20,21] passivation of Pb surface states is crucial for the fabrication of efficient devices. PbS QDs are initially kept stable in suspension by means of long-chained oleate ligands bound to their lead terminated (111) facets.[22,23] While these oleate molecules ensure colloidal stability, their electronically insulating nature makes it necessary to substitute them with shorter-chained ligands prior to their application in optoelectronic devices.[24,25] The most common strategy for such a ligand exchange in PbS QD solar cells is based on the use of lead halide salts ($PbX_2$, X = I or Br), which replace the colloid-forming oleate ligands in the liquid phase.[26] These iodine-rich $PbX_2$ ligands are of anionic nature and bind preferentially to the lead-rich (111) facets of the PbS QDs, in a similar manner to oleate ligands.[27] While ligand exchange with $PbX_2$ significantly improves the efficiency of charge transport in PbS QD thin films, it offers little to no control over the interdot spacing, which, in addition to bound ligands, mostly consists of residual amorphous lead halide species. This is evidenced by the fact that the amount of the Pb halide matrix around PbS QDs has been reported to be as high as 40-50% relative to the PbS.[26,28] Furthermore, the butylamine solvent employed for QD deposition is known to etch Pb halides from the surface of PbS dots, potentially increasing the amount of residual lead halides around the dots.[29] Several strategies for managing this interdot spacing have been proposed, both by chemical and physical approaches. Physically, the effect of temperature and annealing of PbS QD films has been studied thoroughly and it has been shown that annealing processes can improve the coupling between dots.[30] The interdot spacing can be reduced as well by placing the PbS QD films under a solar simulator, but this also triggers a degradation of the electronic transport abilities by generating isolated clusters of densely-packed QDs.[31] Altering the coupling between PbS QDs can be achieved in a chemical manner by modifying the ligands during or after ligand exchange. The chain length of amines used for PbS-$PbX_2$ ink layer deposition can be optimized to yield a dense QD matrix with increased charge diffusion length.[26] Introducing very short ligands, such as HI, will reduce the interdot spacing and improve the passivation on the surface, while still keeping electronic confinement effects.[32] Degradation effects will also alter the spacing between dots. While oxidation causes dot cores to become more separate,[33] humid environments tend to fuse dots under heating.[27] A recent work by Sun *et al*. who have soaked the PbS QD layers in a formamidinium bromide (FABr) containing solution.[34] The authors have shown that during this soaking procedure, the lead halide ligands are dissolved from the PbS QDs and react with the FABr to form a three dimensional (3D) perovskite matrix that bridges between neighboring dots. Such perovskite interdot bridges increase the diffusion length of charges, as well as the efficiency and stability of solar cells. Similarly, a few studies exist exploring the possibility to control the interdot spacing prior to the deposition of the film. For example, a recent study by Li *et al.* showed that the incorporation of tetrabutylammonium iodide (TBAI) during the PbS-PbI$_2$ one-step synthesis can modify the interdot lead halide matrix composition.[35] The authors claim that the addition of these ammonium salts lowers the excess PbI$_2$ between neighboring QDs and results in improved performance of the fabricated solar cells. These examples demonstrate the increased interest in controlling the interdot spacing in QD solar cells as a means for further improvements in efficiency and stability.

In this study, we demonstrate a novel post-fabrication treatment approach for interdot space management in PbS QD solar cells. Rather than replacing the lead halide ligands with a perovskite matrix, we demonstrate that it is possible to manage the excess of lead halides by the use of organic iodide salts, typically employed for the formation of two-dimensional (2D) perovskites.[36,37] Phenethylammonium iodide salts have been used as an additive for PbS QD solar cells as reported recently by Yang et al.[38] The authors claim that the incorporation of small amounts (1 mg/ml) of these molecules enables the passivation of the PbS quantum dots. We, on the other hand, demonstrate that by treating the PbS QD layer with a solution containing either 2-phenylethylammonium iodide (PEAI), 2-(4-fluorophenyl)ethylamine (F-PEAI), or 2-(4-chlorophenyl)ethylamine (Cl-PEAI) (see **Figure S1** for chemical structures), the interdot excess lead halides can be managed, and that no X-PEAI salts remain in the bulk of the PbS film. This occurs via a reaction between these salts and the excess lead halides to form 2D perovskite phases and crystalline PbX$_2$ domains. Selectively dissolving these 2D perovskites and/or PbX$_2$ domains results in PbS QD active layers, that when incorporated in in solar cells, result in higher photovoltaic performance and stability. Our results reveal that the choice of organic cation influences the efficacy of its interaction with the excess lead halides, thus providing important insights for the future application of 2D perovskite cations for interdot space management in PbS QD solar cells.

## 2. Results

The devices under study were fabricated using a common recipe for PbS QD solar cells that was first reported by Liu *et al.* in 2016.[39] In short, PbS QDs were synthesized following a previously reported procedure and exhibited an absorption peak at ~950 nm,[40,41] which corresponds to a diameter of approximately 3 nm. Lead halide salts ($PbI_2$ and $PbBr_2$) were mixed and used as ligands (in the following referred to as $PbX_2$) with an iodine-rich molar ratio of 10:4.[39] Following the ligand exchange, the PbS QD ink was deposited on the solar cell substrates by spin-coating and annealed at 75° C. Next, a solution of X-PEAI cations in acetonitrile (MeCN) was spin-coated on top of the PbS QD films and the samples were annealed at 75 °C to form the 2D perovskite. Finally, the films were rinsed with neat MeCN to dissolve the formed 2D perovskite. These steps are schematically summarized in **Figure 1.** For comparison, reference samples underwent the identical procedure, yet no X-PEAI salt was employed in the first MeCN step so that the devices were simply treated with neat solvent in both steps.

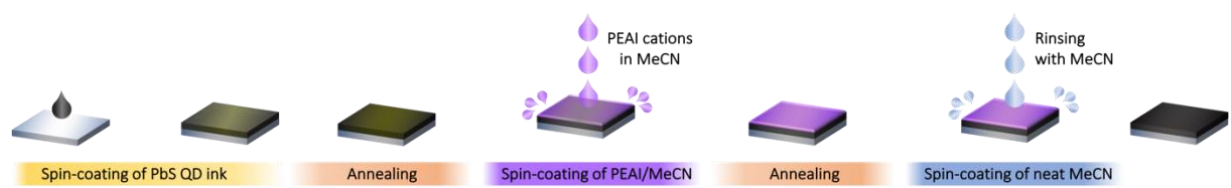

**Figure 1.** Schematic summary of the fabrication steps of PbS-$PbX_2$ active layers. The spin coating of the PbS ink is followed by annealing at 75 °C. Then the X-PEAI salts are dispensed

on top of the PbS QD layer and spun to dry. After an annealing step at 75 °C to form the 2D perovskite, the films are rinsed with neat acetonitrile and spun until dry.

Both the reference and the X-PEAI salt treated PbS QD films were strongly light-absorbent, dark and glossy. It is worth to note here that the reference samples were always rinsed twice with MeCN in order to be more comparable to the treated ones. UV-vis absorption spectra (**Figure 2**a) showed that the first excitonic peak is located at 975 nm and is unaffected by the X-PEAI salt treatment. There is also no indication of agglomeration or increase of the effective QD size, since the peak is not shifted into the IR. The emission spectra of the various PbS QD films are in the IR (at ~1100 nm), as expected due to the Stokes shift present in PbS QDs.[42] Figure 2b shows that the X-PEAI treated films result in a slightly red-shifted emission and a small decrease in the photoluminescence (PL) intensity as compared to the reference PbS QD samples. This is particularly evident following the treatment with the two halogenated X-PEAI derivatives. Both of these effects are suggestive of an improved interdot coupling and a faster exciton dissociation,[43] suggesting that the X-PEAI salt treatment influences the interdot spacing.

To investigate further the influence of treatment with X-PEAI salts on the PbS QD films, we employed Fourier transform infrared (FTIR) spectroscopy and X-ray diffraction (XRD). We note that FTIR measurements on spin-coated PbS layers were too noisy due to their relatively low thickness, so the FTIR measurements shown in Figure 2c and 2d were measured on drop-cast PbS films. Directly after the treatment with the X-PEAI salts (Figure 2c), two bands originating from the primary amine vibrations can be observed: a N-H stretch centered at 3200 cm$^{-1}$, and an N-H bend at 1600 cm$^{-1}$. The first is observed for all of the samples, whereas the second only appears once the PbS has been treated with the X-PEAI salts. The N-H stretch is associated with the presence of the butylamine solvent used in the fabrication of the PbS films since it binds to PbI$_2$ molecules through the amine group.[39] We note that this residual solvent is probably related to the fact that the films were drop-cast, and is likely to be significantly reduced in spin-coated samples. The additional ammonium group from the X-PEAI salts lead to the signal at 1600 cm$^{-1}$ (N-H bend), which is also observed in FTIR measurements of the powder of PEAI and its halogenated derivatives (**Figure S1**). This molecular vibration is difficult to distinguish from residual butylamine at the surface of the QDs after deposition, which can form a complex with PbI$_2$ and PbI$^+$ cations.[29] Consequently, a small contribution can also be seen for untreated PbX$_2$-PbS films. Another feature of the X-PEAI-treated films appears in the region around 1500 cm$^{-1}$. Here, the aromatic carbons from the phenyl ring create distinct vibrational signatures. These are especially prominent for F-PEAI and Cl-PEAI, where distinct peaks appear (see Figure S1 for comparison). Upon rinsing the films with MeCN, the features associated with the X-PEAI salts are greatly reduced (Figure 2d), suggesting the vast majority of the X-PEAI salts are removed from the PbS QD samples.

X-ray diffractograms measured on spin-coated PbS QD films treated with X-PEAI salts are shown in Figure 2e. While the reference, untreated films exhibit only the reflections associated with PbS, the X-PEAI-treated films display additional features. All three treated samples show reflections associated with the 2D perovskite (X-PEA)$_2$Pb(Hal)$_4$ (X=H, F, Cl; Hal = Br, I). All X-PEAI treated samples exhibit more than one distinct peak around 5°. We believe that the presence of iodide and bromide results in the formation of 2D-perovskite with varying mixed halide composition. The reflections of the 2D-perovskites using F-PEAI and Cl-PEAI are generally shifted to smaller diffraction angles when compared to the PEA-treated samples suggesting an increased lattice spacing. This could either be caused by the slightly larger length of the halogenated PEA cation or an overall higher iodide content of the perovskites formed with F-PEAI and Cl-PEAI.[44] In the case of treatment with PEAI, only the main (001) reflex of the 2D perovskites is observed, as well as the (001) reflection of phase-segregated PbI$_2$ (2θ = 12.6°). The treatment with the halogenated F-PEAI and Cl-PEAI results in more intense reflections from the corresponding 2D perovskites and entire Bragg series (001) can be observed. Films treated with Cl-PEAI exhibit phase-separated $PbBr_2$ domains, evident from the corresponding (002) and (111) reflections in the diffractogram.

Upon rinsing the films with MeCN (Figure 2f) the reflections associated with the 2D perovskite can no longer be observed, suggesting this crystalline material was effectively washed off. Similarly, the reflections associated with $PbBr_2$ are also removed, while those arising from $PbI_2$ in the PEAI treated PbS films remain. The origin of these differences will be discussed in detail at a later point.

Due to the low intensity of the XRD signals (mind the log-scale in Figure 2e and f), we repeated the experiments on thicker, drop-cast PbS QD films (**Figure S2**). These experiments confirmed that treatment with PEAI leads to the formation of a 2D perovskite (evident by the presence of entire (001) Bragg series), with still lower intensities compared to the corresponding reflections in F-PEAI and Cl-PEAI treated films. Similar to the results on spin-coated films, only PEAI leads to the formation of phase-separated $PbI_2$, while phase separated $PbBr_2$ appears in the case of the halogenated X-PEAI salts. Upon rinsing these thick, drop-cast films with MeCN, the samples still exhibit small, measurable contributions from 2D perovskite, probably due to the reduced efficacy of MeCN washing away such large amounts of 2D perovskite from the much rougher surface of the drop casted samples. Similarly to the spin-coated films, however, the $PbI_2$ of the PEAI-treated samples remained, while the $PbBr_2$ can be easily removed by acetonitrile.

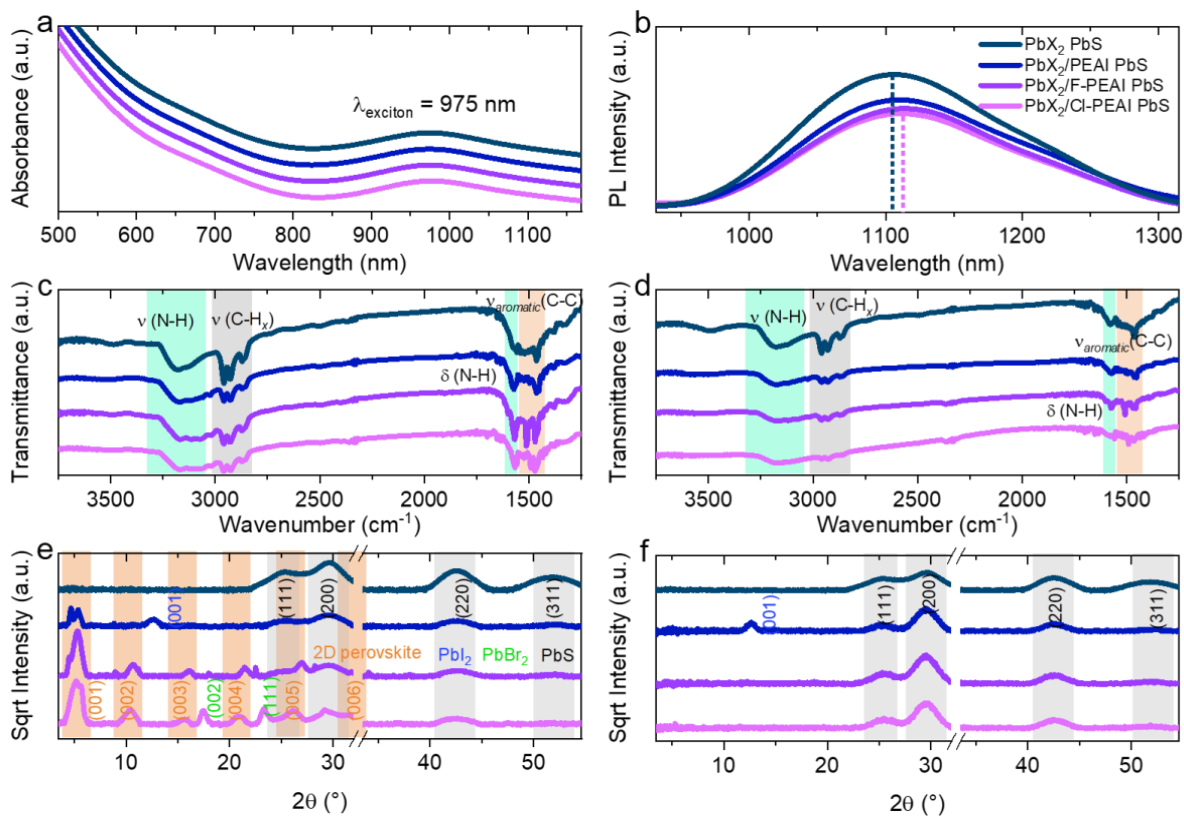

**Figure 2.** (a) UV-vis absorption and (b) photoluminescence spectra of PbS QD films upon treatment with X-PEAI salts. FTIR spectra of drop-cast films (c) before and (d) after the MeCN rinse. XRD measurements on spin-coated X-PEAI salt treated PbS QD films (e) before and (f) after the MeCN rinse.

To further investigate the changes to the composition of the PbS QD films upon treatment with the X-PEAI salts, the samples were measured *via* X-ray photoemission spectroscopy (XPS), both before and after rinsing with MeCN. The I 3d, Br 3d and N 1s spectra are presented in **Figure 3**, while the Pb 4f, F 1s and Cl 2p spectra are shown in **Figure S3**. Upon treatment with X-PEAI salts, the iodine content in the samples is increased (Figure 3a), which is to be expected since all X-PEAI salts contain iodide. Interestingly, the bromine signal is significantly reduced (Figure 3b), in particular for the PEAI-treated sample. The increased iodine content in the case of PEAI, even after MeCN rinse, is a result of the formation of phase separated crystalline $PbI_2$ on the surface, as observed by XRD and confirmed by SEM. On the other hand, treatment with the two halogenated X-PEAI salts shows almost no traces of phase-separated lead halides after the rinsing step. The N1s spectra reveal the presence of two different species: one at 399.5 eV, associated with residual solvent butylamine and one at 402.5 eV, that originates from the nitrogen in the X-PEAI salts. The differences in the intensity of the high binding energy species suggest that different amounts of X-PEAI salts remain in the PbS QD films after treatment. The highest amount is observed for the F-PEAI treated sample, followed by that of Cl-PEAI. This is in agreement with the atomic percentages of fluorine and chlorine (extracted from the F 1s and Cl 2p spectra, Figure S3b and S3c), which were found to be 6% and 3%, respectively. Upon rinsing, the excess of iodine in the treated samples is greatly reduced (Figure 3d), however still surpassing that of the reference sample while the bromine content (Figure 3e) remains significantly reduced. The increase in the iodine content and the reduction in the bromine content suggests that a halide exchange processes occurs during the treatment with the X-PEAI salts. The N1s spectra (Figure 3f) reveal that the intensity of the high binding energy nitrogen species is reduced upon rinsing with MeCN, but is not entirely eliminated from the samples in the washing step. This is supported by the F 1s and Cl 2p spectra, that after rinsing with MeCN, still show clear signals (constituting ~2 atomic %) for each of the halogenated X-PEAI treated samples. Calculating the I/Pb and Br/Pb and S/Pb ratios (Figure 3g, 3h and 3i, respectively) of the samples after rinsing visualizes the changes in composition that are induced by the X-PEAI treatment. The small increase in the S/Pb ratio (by up to 5%) suggests that the treatment led to the removal of some $PbX_2$ species. S 2p XPS spectra of MeCN-rinsed samples are shown in **Figure S4**. In addition, the increase in I/Pb and decrease in Br/Pb ratios led to a significant modification of the I/Br ratio from the initial level of 10:4 introduced during synthesis to 10:1.3 for PEAI, 10:2 for F-PEAI, and 10:2.5 for Cl-PEAI. The fact that the changes in the halide content are far larger than those in the lead content confirm that a halide ion exchange process accompanies the removal of the excessive interdot $PbX_2$. To investigate if the changes are confined only to the surface of the films or if they impact also its bulk, we performed XPS depth profiling experiments (**Figure S5**). The experiments revealed that while the surface is particularly halide rich, the bulk is also impacted by the treatment with X-PEAI. Specifically, upon treatment, the I/Pb ratio is increased, while the Br/Pb ration is decreased in the bulk of the samples. This confirms that the removal of excess lead halides from the bulk of the sample is accompanied by halide exchange. Moreover, depth profiling experiments revealed that neither F nor Cl signals are present in the bulk of the sample, suggesting that no X-PEAI remains in the bulk of the film (**Figure S6a**). This observation confirms that the treatment with X-PEAI does not lead to the formation of a 2D perovskite shell around the PbS QDs, or a 2D perovskite matrix in between the QDs. The higher presence of halides on the surface of PbS films, even for the reference, favors the formation of 2D perovskite on the surface of the sample. This could be due to the interplay between lead halides and butylamine during the spin-coating and annealing,[29] forming a lead halide rich film surface. Diffusion into the densely packed film by the large PEA$^+$ cations is also unlikely, so the formation of 2D perovskite inside the bulk of the film is not to be expected. This is furthermore confirmed by performing tunneling electron microscopy imaging of Cl-PEAI treated films (**Figure S6b**). The diffraction patterns resulting from a Fourier transform of the images reveal exclusively the features associated with PbS, thus confirming that no 2D perovskite shell remains on the QDs.

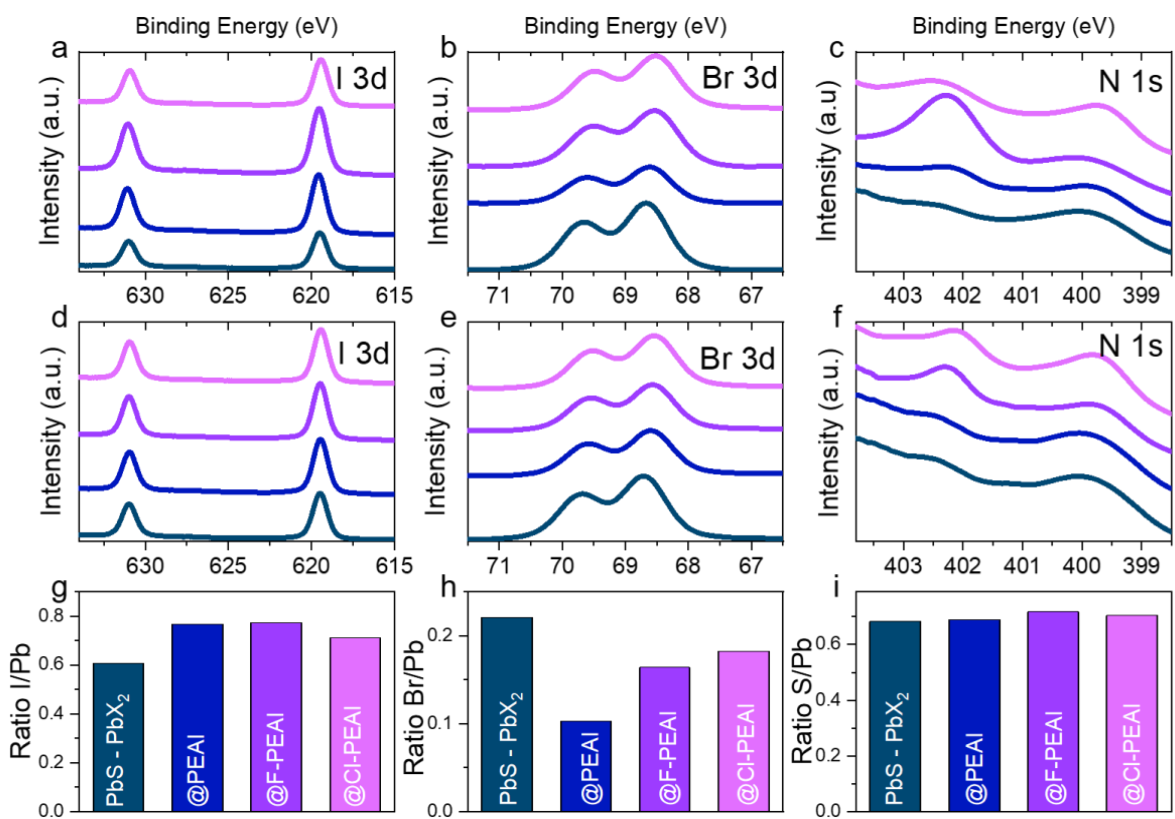

**Figure 3.** (a)/(d) I3d, (b)/(e) Br3d and (c)/(f) N1s XPS spectra of PEAI-cations treated samples before/after MeCN rinse. (g) I/Pb (h) Br/Pb and (i) S/Pb ratios of PEAI after MeCN rinse.

The compositional changes observed *via* XPS suggest that the microstructure of the PbS QD films might also have been altered. To visualize the surface microstructure, all films were further investigated using top-view scanning electron microscopy (SEM), and are shown in **Figure 4**. Reference PbS QD films exhibit a smooth surface and a compact active layer (**Figure S7**). Upon treatment with PEAI, small hexagonally shaped crystallites are formed on top of the layer (Figure 4, left) and are not affected by the step of the MeCN rinse. Based on the results of the XRD and the shape of the crystallites, we associate them with crystalline $PbI_2$. In the case of the halogenated PEAI derivatives, there is a noticeable roughening of the films, which we believe arises from the formation of thin layers and domains of 2D perovskite and in the case of Cl-PEAI, some $PbBr_2$ domains. These surface features are completely removed after the MeCN rinse and the films appear smooth and structurally intact afterwards, similar to state prior to X-PEAI application.

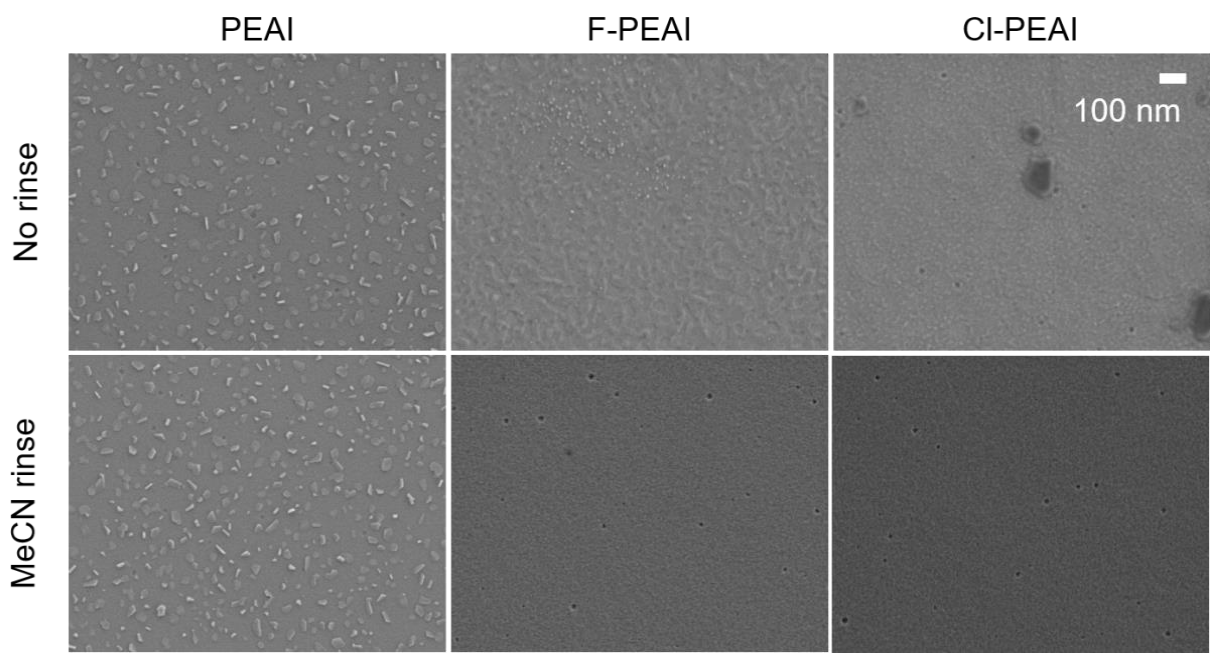

**Figure 4.** Top view SEM images acquired with Inlens detector of PbS QD films before (top) and after (bottom) treatment with X-PEAI salts and subsequent MeCN rinse. Identical scale bar for all images, given at the top right.

To investigate the impact of the X-PEAI treatment on the photovoltaic performance of the QD devices, the PbS QD active layers have been integrated into solar cells with the structure glass/ITO/ZnO/PbX$_2$-PbS/EDT-PbS/Au. Figure 5a-d displays the photovoltaic performance of devices treated with different concentrations of X-PEAI salts normalized to that of reference, untreated devices. Importantly, all of these devices have been fabricated using the same PbS ink and identical processing conditions in one very large batch and thus allow for a clear evaluation of the impact of the treatment with X-PEAI of different concentrations. It can be seen that the treatment leads to a small increase in the V$_{OC}$, especially for the low concentration, but a much more pronounced increased in the J$_{SC}$ and FF. Overall the optimum PCE is achieved for a concentration of 5mg/ml of X-PEAI and is substantially higher for the Cl-PEAI treated devices, that show an improvement by 10-15%. The increase in $J_{SC}$ is evidenced by a higher external quantum efficiency (EQE) spectra of the treated devices (**Figure S8a**), that not only lead to higher overall response, but also show a small shift in the position of the first excitonic peak towards lower energies upon treatment (~10 nm redshift, as observed in PL measurements), which stems from a better interdot coupling similar to the one observed by researchers after FABr treatment.[28] A redshift of the EQE interference peak at ~750 nm is observed for X-PEAI treated devices, is likely associated with an increase in the effective refractive index of the film due to the more dense packing of the quantum dots following the X-PEAI treatment. Similar effects have been observed for textured PbS QD devices in the past.[45,46] *J-V* curves of selected pixels under one sun illumination are combined in **Figure S8b**. The comparison of the PCE of multiple devices of each kind from different batches is shown in **Figure 5e** with all photovoltaic parameters shown in **Figure S9**. The highest performance is obtained for the Cl-PEAI treated devices, who reach a maximum PCE of 11.8%. While these data show a broader distribution in photovoltaic performance since they originate from different batches, they clearly demonstrate that the treated devices result in a higher power conversion efficiency.

To gain insight into the origin of the improved performance the devices were characterized under different illumination intensities (**Figure S10** and **Figure S11**). The light intensity-dependent $V_{OC}$ plots show that X-PEAI treated devices show a decreased $kT\ q^{-1}$ slope, which is associated with a reduced trap-assisted recombination.[47] The current vs. intensity data suggest only minor differences as compared to the untreated reference solar cells. Furthermore, in order to more directly probe any potential changes to the defects states in the PbS quantum dot solar cells, we applied pump-push-photocurrent (PPPC) spectroscopy which enables us to probe the density and population dynamics of trapped charges in the devices (**Figure S12**).[48] The PPPC experiment is performed on functional solar cells at working conditions and the measured signal is proportional to the concentration of immobile (trapped) carriers at the respective delay times. The decay of the signal is thus assigned to the recombination of the trapped carriers. In agreement with a previous study,[48] charge trapping in the PbS devices occurred on ns timescales and reached its maximum at ~300 ns. This is followed by ~10 ms recombination. We find that the X-PEAI treatment does not result in the introduction of additional defects, as both the concentration and dynamics of the trapped charges are very similar in both the reference and the treated samples.

The performance of fabricated solar cells was studied over time and characterized as shelf-storage lifetime. The devices were kept unencapsulated in air (~30-40% relative humidity, 23 °C) under dark conditions and measured with the solar simulator in diverse time intervals. **Figure 5f** shows an increased stability of devices treated by X-PEAI salts. The PCE of the F-PEAI and Cl-PEAI cells maintains 95% of their initial value after 500 hours of storage, while the PEAI treated devices drop to 87% of the initial PCE. In comparison, the PCE of the $PbX_2$-PbS reference devices dropped to 75% of the initial value within the same time interval. We note that the long-term dark storage stability of our devices is lower than the that observed in other reports for devices,[39] which we attribute to differences in the processing of the devices and the environment (especially relative humidity).[46] Longer-timescale degradation is dominated mostly by degradation of the hole transport layer (PbS-EDT),[33,49] and is therefore not impacted by the X-PEAI treatment.

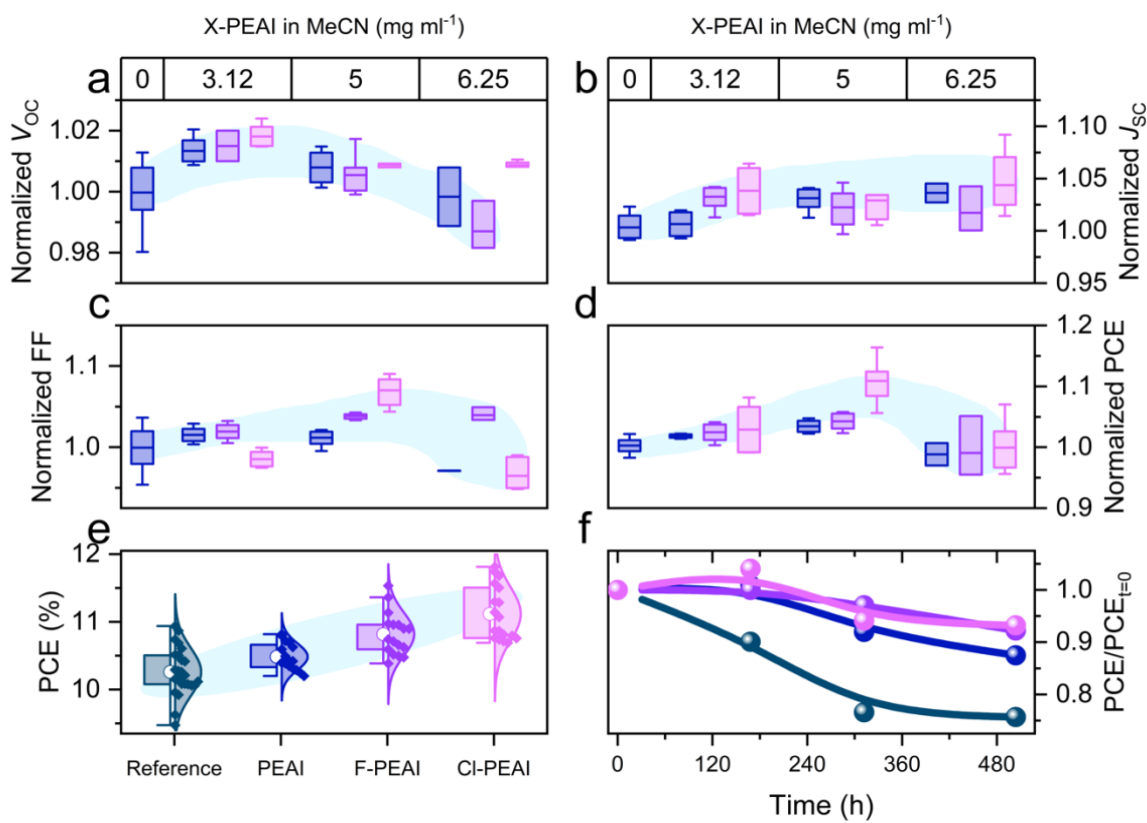

**Figure 5**. (a) Open-circuit voltage, (b) short circuit current density, (c) fill factor and (d) power conversion efficiency of PbS QD photovoltaic devices as a function of X-PEAI salt concentration in MeCN as top treatment. (e) The overall distribution of PCE shows increased performance for all treated devices. (f) Lifetime curves of reference and treated devices after dark storage in (30-40 % relative humid) air.

The results of the spectroscopic and microscopic characterisations and the improved photovoltaic performance and stability of the solar cells suggests that the treatment with the X-PEAI salts allows to manage the interdot lead halide excess as well effectively modify the surface composition of the solar cells' active layer. To confirm the improved interdot coupling in treated devices, the hole mobility was measured in unipolar device structures. For this purpose, hole-only devices consisting of ITO/PbS-EDT/PbS-PbX$_2$ (150 nm)/PbS-EDT/Au were fabricated for both reference and X-PEAI-treated devices. The mobilities extracted by the space charge limited current method ($J \sim V^2$), shown in **Figure S13**, are a factor of two higher for F-PEAI and Cl-PEAI treated devices, corroborating that this treatment leads to improved interdot coupling. To further confirm the changes in the interdot spacing and the resulting improved interdot coupling, we performed TEM measurements in which the ligand exchange and subsequent treatment with X-PEAI was performed directly on the TEM grid, following the procedure reported by Teh et al. to quantify the interdot spacing after a thiol ligand exchange.[50] In the exemplary case of F-PEAI, the interdot spacing is reduced from 1.2 nm to 0.8 nm (**Figure S14**). We highlight that this dot-to-dot distance is larger than the one in the solid state PbS-PbX$_2$ films as measured by Grazing-Incidence Small-Angle X-ray Scattering (GISAXS), which is around 1-2 Å,[28] and stems from the deposition technique on TEM grids from heavily diluted inks. This decrease leads to improved interdot coupling, which is evidenced by a reduction in the PL lifetime of treated samples probed by time correlated single photon counting (TCSPC) experiments (**Figure S15**). The TCSPC measurements show that the lifetime of PbX$_2$-PbS reference samples is reduced from 2.3 ns to 1.8 ns upon treatment with F-PEAI. This decrease is related to the improved exciton dissociation in quantum dot films with reduced interdot spacing.

Since treatment of the quantum dot active layer with pure MeCN in the reference devices does not lead to such modifications, the presence of the PEAI components in the MeCN must facilitate the changes in layer and surface composition (**Figure 6**). Upon the addition of the X-PEAI salts in the first MeCN step, followed by a gentle drying at 75°C, 2D-perovskite domains are formed on top of the active layer. We believe that the presence of X-PEA$^+$ and I$^-$ ions increases the chemical mobility of lead halides towards the surface of the porous quantum dot film, leading to an effective extraction of lead halides from the active layer itself. It is fair to assume that lead halide aggregation, crystallisation and perovskite formation and the interaction with X-PEAI act as the thermodynamic driving force for this process. We emphasize, that the concentration and therefore overall amount of the X-PEAI salt is crucial and needs to be finely adjusted (**Figure 5a-d**). An effective removal of the lead halide excess and best PV performance was found for a concentration of 5mg/mL in the case of Cl-PEAI. Higher concentrations lead to a degradation of the device efficiency and could indicate that not only excessive lead halides are being removed from the film but also the $PbX_2$ ligand shell around the quantum dots might be affected or even etched when higher amounts of X-PEAI salts are being used.

Since lead iodide and lead bromide are present in the active layer, we expect that the overall amount and ratio of phase separated lead iodide, lead bromide and perovskites domains, as well as the halide composition of the 2D-perovskite itself, strongly depend on the interaction of the interdot lead halides with the X-PEAI salt. In the case of PEAI, we observed phase separated $PbI_2$ domains, which suggests a higher bromide content in the 2D-perovskite domains. While for Cl-PEAI and F-PEAI the opposite effect occurs and phase separated lead bromide is formed. When washing the films after X-PEAI treatment with MeCN, only the lead iodide domains remain while the lead bromide ones are washed off. This can easily be explained by the ca. 10 times higher solubility of pure lead bromide in MeCN compared to lead iodide (see **Figure S16**). The formed 2D perovskites exhibit a sufficiently large solubility in the applied volume of MeCN and are removed from the film's surface in all three cases and only very minor traces (ca. 2%) of X-PEAI remain. To further quantify the excess Pb halide removal from the film, we calculated the weight loss after drop casting treatment and a subsequent MeCN rinse (**Figure S17**).

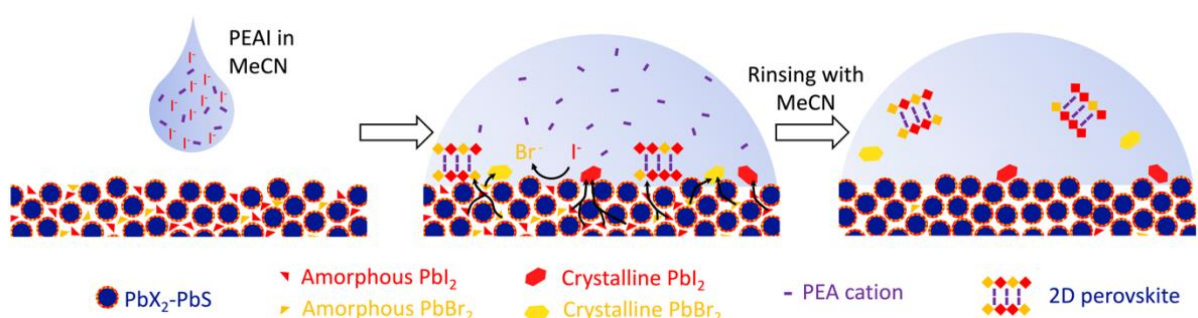

**Figure 6.** Schematic presentation of the processes that occur during the PEAI cation modification of PbS QDs.

Considering the suggested mechanism for excess lead halide removal above, it is important to notice that all steps involve the interplay and competition of two different anions. While the employed X-PEAI salts are all iodide based, the excess of lead halide $PbX_2$ in the quantum dot film contains bromide and iodide. That in turn means that exchange processes between iodide and bromide will occur and might even reach an equilibrium state within the application time of MeCN solution. Generally, the use of iodide-based salts in the solvent will cause in all cases a shift of the I/Br ratio to higher iodine contents. However, the exact ratio will depend on the differences in solubility, crystallisation dynamics and formation enthalpy between the corresponding bromide and iodide species. For example, the overall ratio of solvated bromide and iodide in the first MeCN step that will be removed by spinning off the MeCN solution after a contact time of 20 sec will most likely depend on the differences in solubility of the corresponding X-PEAI and X-PEABr salts. Our experiments indicate that PEAI alters the initial I/Br ratio more severely, while F-PEAI causes the smallest change in the halide ratio. The underlying reason to a higher device performance with F-PEAI and Cl-PEAI in our study is related to the fact that these molecules have a higher dipole moment than pure PEAI,[38] thus triggering a different effect in excess removal and later solubility of products on the surface.

## 3. Conclusion

This work presents a facile approach to increase the efficiency and shelf life of PbS QD solar cells by 2D perovskite cation mediated removal of excess interdot lead halides. We demonstrate that upon the addition of small concentrations of PEAI, Cl-PEAI or F-PEAI salts, excess $PbI_2$ and $PbBr_2$ between the quantum dots can be extracted and transformed into 2D perovskites or crystalline $PbX_2$ on the surface of the QD films. These can be subsequently rinsed, leading to a film with increased interdot coupling, an improved photovoltaic performance and enhanced stability. Our work opens a new route for an effective management of excess lead halides in lead chalcogenide solar cells and other optoelectronic devices.

## 4. Experimental Section/Methods

*Materials and purity:* Lead (II) oxide (Alfa Aesar Puratronic, 99.999%), Oleic Acid (Alfa Aesar, 90 % technical grade), Octadecene (Alfa Aesar, 90 % technical grade), Hexamethyldisilathiane (Sigma Aldrich, synthesis grade) are employed as received for PbS QD synthesis. Lead (II) iodide and lead (II) bromide (TCI, 99.99% trace metal basis), Ammonium acetate (Fisher Chemical, >97%) and *N,N*-Dimethylformamide (Acros Organics, 99.8% ExtraDry) are used in the ligand exchange. PEAI, F-PEAI and Cl-PEAI were purchased from Greatcell Solar Materials and have a purity of >99%. They should be stored in a nitrogen glovebox since they will absorb water. All organic solvents used in this study are from Acros Organics.

*PbS QD synthesis*: For a usual batch of QDs the hot-injection synthesis recipe was adapted from previous works.[40] In order to get PbS QDs with a bandgap of 1.3 eV, PbO (451 mg), OA (1.5 ml) and ODE (18 ml) were degassed in a $10^{-3}$ mbar vacuum under 100° C using Schlenk techniques. After 2h, TMS (180 μl) was swiftly injected into the clear solution under nitrogen. Instantly the nucleation started, and the color changed to dark brown. The reacted product was collected and purified using acetone (1:1 by volume) and methanol/butanol (1:8 by volume), always collecting the QDs by a 10-minute centrifugation at 5000 rpm after adding the antisolvents. For smaller dots in the PbS-EDT layer, the recipe is the same except for the injection temperature, which is reduced to 90° C to get smaller dots.

*PbS-PbX$_2$ ligand exchange:* For a typical ligand exchange, a mixed lead halide precursor is prepared in N, N-Dimethylformamide (DMF). PbI$_2$ (0.1 M), PbBr$_2$ (0.04 M) and ammonium acetate (0.04 M) are dissolved in DMF (20 ml). This yellow precursor is added to PbS (20 ml) QDs in octane (10 mg/ml). After a vigorous mixing, the black DMF phase is collected and washed three times with pure *n*-octane. The remaining liquid is then placed in a centrifuge tube and after adding toluene (1:0.75 by volume) it is centrifuged for 5 minutes at 5000 rpm. The supernatant is discarded, and the remaining pellet is dried under vacuum for an hour. The pellet (normally around 150 mg) can be stored in an inert atmosphere for at least a week before making the solar cells.

*Device fabrication:* Pre-patterned glass/indium tin oxide (ITO) substrates (12 x 12 x 1mm) are cleaned using acetone and isopropanol in a sonication bath. The surface is activated with oxygen plasma for 10 minutes. A thin ZnO sol gel layer is spun at 3000 rpm on the substrate and annealed at 300° C, following published methods.[51] The stored PbS-PbX$_2$ pellet is dissolved in *n*-butylamine with a concentration of 400 mg/ml, depending on the desired thickness. This dense ink is spin coated on the substrates for 20s at 2500 rpm and dried on a hotplate for 10 minutes at 75° C. This step is performed in a nitrogen glovebox to ensure reproducible results. For solar cells with additional treatment, a solution of either PEAI, F-PEAI or Cl-PEAI (5 mg ml$^{-1}$) is prepared with dry acetonitrile (MeCN). This solution is dropped on top of the active layer and spun after 20 s at 3000 rpm to remove excess solvent. The substrates are placed again on a plate at 75° C to form the 2D perovskite (10 minutes) and then they are rinsed with neat MeCN on the spin coater. Two layers of PbS-EDT are spin-coated in air using already published methods.[40] It is important to note that we used PbS dots with a first excitonic peak around 850 nm for this layer, to provide a better blocking of electrons in the solar cells. 80 nm thick Au pixels are thermally evaporated (10$^{-6}$ mbar) on top to give 4.5 mm$^2$ solar cells. There is a total of 8 solar cells per substrate.

*Absorbance and luminescence*: The absorbance and photoluminescence (PL) of the PbS QDs is measured on films made on glass by spin-coating. The absorbance spectra are measured with a Shimadzu SolidSpec 3700. The PL signal is acquired using a green LED (Thorlabs, M505L3) as excitation source and an NIR InGaAs spectrometer (Instrument Systems, CAS 140CT IR1) as detection unit. To assess the exciton lifetime of the quantum dots, QD films with a photoluminescence peak centered around 950 nm were investigated. Corresponding TCSPC measurements were conducted in air utilizing a pulsed 405 nm laser (LDH-IB diode laser and Taiko PDL M1 controller, both PicoQuant GmbH) as excitation source. To ensure only the emitted photons from the QD layer are counted, an 850 nm longpass filter (Thorlabs) was used. The measurements utilized a fibre coupled PDM SPAD (Micro Photon Devices) and the PicoHarp 300 electronics (PicoQuant GmbH) with an integration time of 96 min. The PL decay was fitted using a convolution of an exponential and Gaussian function.

*Infrared spectroscopy*: Fourier-transformed IR (FT-IR) spectra are taken with a Shimadzu IR Spirit. The samples were drop-cast on single side polished silicon substrates and measured in transmission mode.

*X-ray photoemission spectroscopy (XPS)*: XPS measurements were carried out in an ultrahigh vacuum chamber (ESCALAB 250Xi by Thermo Scientific, base pressure: $2 \times 10^{-10}$ mbar) using an XR6 monochromated Al Kα source (hν = 1486.6 eV) and a pass energy of 20 eV. XPS depth profiling was performed using an argon gas cluster ion beam with large argon clusters ($Ar_{2000}$) and an energy of 4 keV generated by a MAGCIS dual mode ion source. XPS spectra were evaluated using the Avantage software by Thermo Scientific. First, a suitable background was added to each spectrum, then the peaks were fitted using Gaussian-Lorentzian functions. Fits of doublet peaks were restrained regarding spin-orbit splitting, peak area ratio and FWHM. Finally, the software was used to calculate the atomic percentage contribution of each element using the resulting peak areas and their respective atomic sensitivity factors. The fitted spectra shown in Figure 3 are show in **Figure S18** and **Figure S19**.

*X-ray diffraction*: PbS films drop-cast or spun on silicon wafers were analyzed using a Bruker D8 Discover diffractometer equipped with a 1.6 kW Cu X-ray filament (λ = 1.5 Å) and a 0.6 mm slit. The scans are performed using a 1D detector in the θ/2θ mode.

*Scanning electron microscopy (SEM):* Images were acquired on a Zeiss Gemini 600 SEM in high resolution mode. The acceleration voltage was set to 1.5 kV, to avoid strong degradation of the sample due to charging.

*Transmission electron microscopy (TEM):* The TEM images were acquired on a Jeol JEM F200 operated at 200kV in transmission mode and images were taken on a Gatan 4k video camera. Samples were drop-cast from a dilute PbS QD ink directly on TEM carbon/copper grids and dried in a glass oven before mounting into the TEM holder. The ligand exchange was also performed directly on the TEM grid.

*Solar cell characterization*: Solar cells were measured under an Abet A+++ solar simulator, with a simulated AM1.5 spectrum that was calibrated using a Newport reference silicon solar cell (100 mW cm$^{-2}$). The forward-backward voltage sweep (-0.1 to 0.7 V in 0.025 V steps) was done with a Keithley 2450 source measure unit (SMU). The intensity dependent JV curves were acquired by placing diverse optical filters (Thorlabs absorptive ND filter kit) with known optical density on top of the measured solar cells. External quantum efficiency (EQE) spectra were measured with a self-built setup. A halogen lamp passed through a monochromator is used to generate a current in the solar cells, which is measured with a Keithley SMU. This current is corrected for the irradiance of the lamp using a calibrated silicon and InGaAs diode and the efficiency to generate an electron for every irradiated photon is calculated.

*Pump-push-photocurrent (PPPC) measurements*: PPPC measurements were performed using the output of a Picolo (Innolas, $\lambda$ = 1064 nm, $\tau$ ~ 2 ns, pulse energy ~0.25 nJ) laser as the 'pump', and an optical parametric amplifier (TOPAS-Prime, Coherent, $\lambda_c$ = 2073 nm, $\tau$ ~ 100 fs, pulse energy ~2 µJ), seeded by a Ti:sapphire regenerative amplifier (Astrella, Coherent), as the 'push'. Pump and push beams were focused onto a ~400 µm-diameter spot at the samples. The pump reference current (*J*) was read out by a lock-in amplifier (MFLI, Zurich Instruments) at 4 kHz, and the push-induced current (*ΔJ*) was modulated at 717 Hz by an optical chopper. The pump-push delay time was controlled by triggering the pump via a delay generator (SRS DG645, Lambda). Due to the delay generator jitter, the time resolution of the experiment was on the order of 20 ns. After the measurements, data were corrected for bimolecular recombination effects, which were estimated by using the dynamics at negative delay times. Data were normalised by the reference current *J* and the *ΔJ*/*J* values were used for the interpretation of the results.[48]

**Supporting Information**
Supporting Information is available from the Wiley Online Library or from the author.

**Acknowledgements**
The authors acknowledge financial support by the European Research Council (ERC) under the European Union's Horizon 2020 research and innovation program (ERC Grant Agreements No. 714067, ENERGYMAPS). We thank the Deutsche Forschungsgemeinschaft (DFG) for funding the project 'PROCES' (VA 991/2-1). We would also like to acknowledge the Dresden Center for Nanoanalysis for providing the opportunity to perform electron microscopy experiments.

Received: ((will be filled in by the editorial staff))
Revised: ((will be filled in by the editorial staff))
Published online: ((will be filled in by the editorial staff))